# WestDrive X LoopAR: An open-access virtual reality project in Unity for evaluating user interaction methods during TOR.

Farbod N. Nezami[1‡], Maximilian A. Wächter[1‡*], Nora Maleki[1], Philipp Spaniol[1], Lea M. Kühne[1], Anke Haas[1], Johannes M. Pingel[1], Linus Tiemann[1], Frederik Nienhaus[1], Lynn Keller[1], Sabine König[1], Peter König[1,2*], Gordon Pipa[1*]

‡ Shared First Authorship
\* Shared Senior Authorship

Affiliation 1: Institute of Cognitive Science, University of Osnabrück, Osnabrück, Germany, mwaechter@uni-osnabrueck.de

Affiliation 2; Department of Neurophysiology and Pathophysiology, Center of Experimental Medicine, University Medical Center Hamburg-Eppendorf, Hamburg, Germany

Correspondence: mwaechter@uni-osnabrueck.de; Tel.: (+49)-541 / 969-2245

**Abstract:** With the further development of highly automated vehicles, drivers will engage in non-related tasks while being driven. Still, drivers have to take over control when requested by the car. Here the question arises, how potentially distracted drivers get back into the control-loop quickly and safely when the car requests a takeover. To investigate effective human-machine interactions in a mobile, versatile and cost-efficient setup is needed. We developed a virtual reality toolkit for the Unity 3D game engine containing all necessary code and assets to enable fast adaptations to various human-machine interaction experiments, including close monitoring of the subject. The presented project contains all needed functionalities for realistic traffic behavior, cars, and pedestrians, as well as a large, open-source, scriptable, and modular VR environment. It covers roughly 25km², a package of 125 animated pedestrians and numerous vehicles, including motorbikes, trucks, and cars. It also contains all needed nature assets to make it both highly dynamic and realistic. The presented repository contains a C++ library made for LoopAR that enables force feedback for gaming steering wheels as a fully supported component. It also includes All necessary scripts for eye-tracking in the used devices. All main functions are integrated into the graphical user interface of the Unity® Editor or are available as prefab variants to ease the use of the embedded functionalities. The primary purpose of this project is to serve as an open access, cost-efficient toolkit that enables interested researchers to conduct realistic virtual reality research studies without costly and immobile simulators.

**Keywords:** VR research, Out-of-the-loop unfamiliarity (OOTLU) Autonomous driving, Human-machine interaction, Take-over Request (ToR)

## 1. Introduction

What defines the user-friendly design of automated systems has been the subject of scientific discussion for decades [1,2]. Especially in upcoming years, when automated vehicles of SAE automation level 3 and 4 will emerge, the demands on the driver's cognitive system will alter radically, as the role of humans as continuously active decision-makers in vehicles is replaced by automated systems [3,4]. Such systems include the Audi traffic jam pilot [5] or Tesla's full self-driving beta [6]. Experiences from aviation, where automated systems are already widely integrated, clearly state that the safety and reliability of such systems cannot be achieved by optimizing technical components alone [7]. Instead, the reliability of highly automated systems is primarily determined by the driver's cognitive processes, meaning how fast a safe transition to manual drive is possible [8].

The need for a fast and safe transition applies particularly to situations where humans have the task of taking over system control in the event of sensor failures or malfunctions [9,10]. Thus, investigating the fluent integration of the takeover request (ToR) is of crucial importance for the safety of any system with even partially automated driving features [11]. During a ToR, the human driver most likely has to take over control in under 10 seconds, even when not engaged in driving-related activities [12,13,14]. Naturally, an orientation phase follows as the human driver has to assess the traffic situation [14]. Unfortunately, the driver's reaction is often too slow in critical situations, potentially resulting in an accident in a small time frame (<4 sec) before an impact occurs [15,16]. Even in the case of fast reactions within a timeframe under 10 seconds, studies with prolonged driving have shown merely hectic responses by human drivers, which of course did neither improve the reaction time nor the situational outcome [17,18].

In this manuscript, we will present a new toolset for human-machine interaction research apart from typical screen-based simulators. Existing simulators are often based on actual car interior designs. Therefore they offer only limited possibilities for the human-machine interaction (HMI) research [19]. A very similar problem is posed by research in prototype cars in the real world, where realistic accident scenarios are costly and can only be generated to a minimal extent without endangering the test person involved. The project, called LoopAR, provides not only all needed assets and an environment but also all needed code to display information of a takeover request as a freely programmable augmented reality (AR) feature in the windshield. The developed HMI displays the takeover request and highlights critical traffic objects to enable participants to take over more quickly and precisely. It is thought to fundamentally change research of safe and effective communication between car and driver. This is not only beneficial in terms of safety for the passengers but could also increase customer acceptance of highly automated vehicles, since up until now, malfunctions are vital concerns of possible customers [20]. Since the LoopAR project is based on the project WestDrive [21], all code needed and the designed scenes are available in a Github repository. This enables interested researchers to create their own experiments with the WestDrive LoopAR toolkit. In order to fully use the project presented here, only a powerful computer, VR glasses, a simulation steering wheel, and pedals, as well as Unity as a development program, are required.

## 2. Methods and Main Features of LoopAR

The main focus of the presented project is versatility and modularity, which allows the fast adjustment of the environmental and functional objects via prefab and the provided code in the toolkit. Research on interactions between humans and cars is mostly done with stationary simulators. Here, a whole car chassis is used, or only the interior is set inside a multi-screen setup. However, these classical setups are often expensive, and adjustments or graphical improvements of the stimuli used in an experiment are often not possible [22]. In the past years, there has been a significant shift of research towards virtual environments. It is reflected by applications like Cityengine and FUZOR [23, 24] and software for driving environments [25]. Still, experimental designs on the Human-Machine Interaction, in terms of specific car interior adjustments, are not possible yet. Therefore the presented project enables the user to create experimental conditions and stimuli freely. All functionalities that are mentioned in the following are independent and can be adjusted at will.

Additionally, the presented project does not need a specific hardware setup, making it also easily adaptable and future-proof. New components like, e.g., new GPUs and new VR devices, can be easily integrated into the setup. The current requirements only apply to the VR devices used and are not bound to the toolkit.

*2.1. Platform*

Project LoopAR is made with the Unity Editor 2019. 3. 0f3 (64bit). This software is a widely used game engine platform based on C# by Unity Technologies, supporting 2D, 3D, AR & VR applications. The Unity editor and the Unity Hub run on Windows and Mac and Linux (Ubuntu and CentOS), and built applications can be run on nearly all commercially usable platforms and devices. Unity also provides many available application programming interfaces and is compatible with numerous VR and AR devices [26].

The backend code of the project LoopAR was developed entirely using C# within Unity3D Monobehaviour scripting API. The backend comprises functionalities including dynamic loading of the environment, AI car controls, pedestrian controls, event controls, car windshields augmented reality controller, and data serialization and eye-tracking connection. Additionally, the presented project contains a C++ library enabling the force feedback for Microsoft DirectX devices that enables various force feedback steering wheels to function as controllers altogether. LoopAR code has been developed with modularity in mind to avoid complicated and convoluted code. All functionalities can be enabled or disabled individually using UnityEditor's graphical interface based on need.

*2.2. Virtual Environment*

In order to test human-machine interaction, an interactive and realistic 3D environment is needed. LoopAR aims at a fully immersive experience of a highly automated car encountering critical traffic events. To be able to investigate different driving conditions and scenarios, we created four independent scenes. In the following section, the environment design decisions will be presented together with a short description of the experimental scenes.

The LoopAR environment is based on real geographical information of the city of Baulmes in the Swiss Alps. The region was selected due to its variety in the terrain, including a small village, a country road, a mountain pass, and a region suitable for adding a highway section, totaling around 25km² of environment and an 11km continuous drive different roads. To reduce computational demands, we sliced the terrain into four areas. Due to the road network design, these separate environments can be merged. (see Fig.1). These areas demand different driving skills to an automated driving vehicle and a human driver to react in different situations with different conditions according to the landscape and traffic rules. To make the region accessible in Unity, we used the collaborative project OpenStreetMap (OSM) [27] and the open-source 3D software Blender

[28].

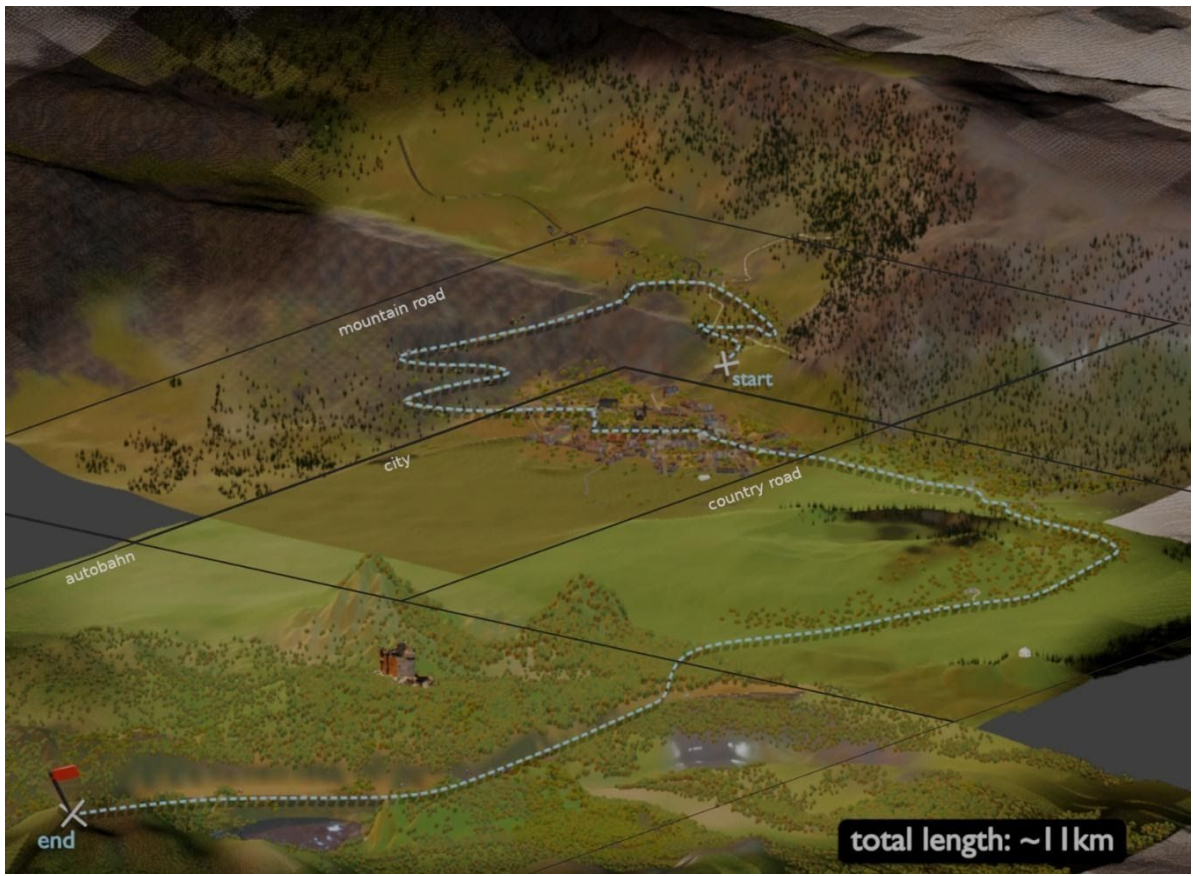

Fig.1: LoopAR map preview: mountain road 3,4km, city 1,2km, country road 2,4km, highway 3,6km.

The virtual environment contains a mountain road scene (see Fig. 2a), including curvy roads winding through a forest and steep serpentines running down a mountain. These curvy roads require various driving speeds (from 30km/h or slower, up to 100km/h on straight stretches). The overall traffic density is low.

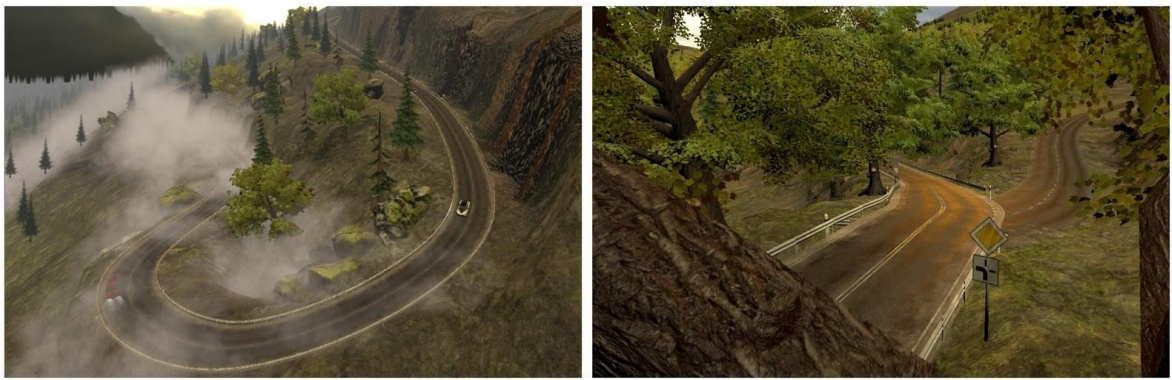

Fig.2a: Pictures of the different scenes: Mountain Road

The second area of the environment is the village "Westbrück" (See Fig. 2b). Here it is possible to test events in a more inhabited environment. This environment is characterized by narrow streets and dense traffic in low-speed environments.

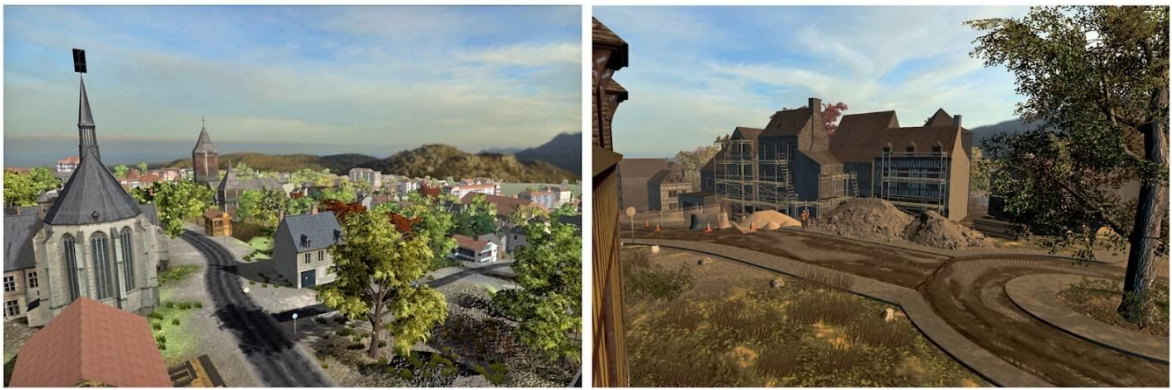

Fig.2b: Pictures of the different scenes: Village "Westbrück."

The third scenario is the country road scene (see Fig. 2c), designed for medium to high speed (~70km/h), medium traffic density, and a long view distance.

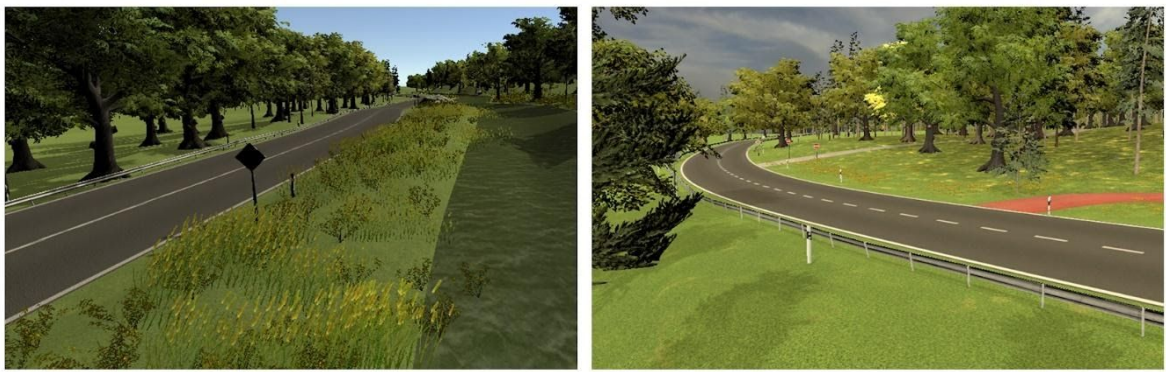

Fig.2c: Pictures of the different scenes: Country Road

The last scenario for the participants is the highway scene (see Fig. 2d), enabling critical traffic events with a higher speed and a low to medium traffic density.

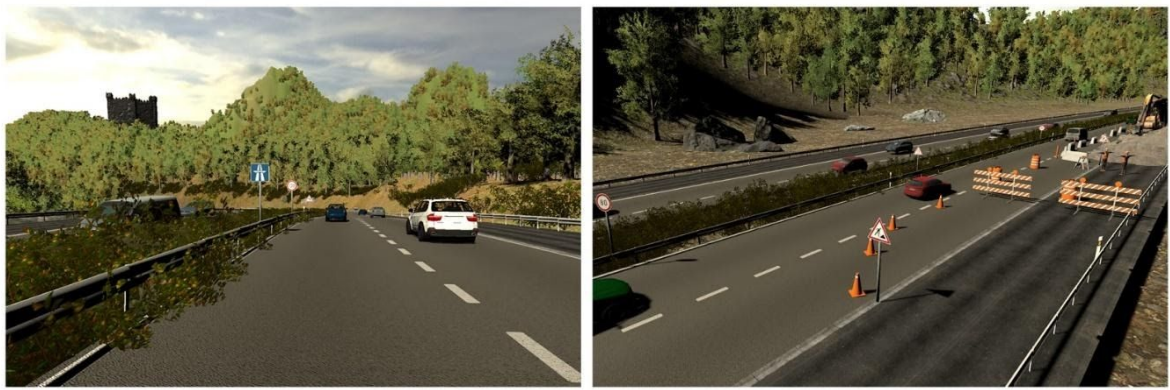

Fig.2d: Pictures of the different scenes: Highway

### 2.3. Critical Traffic Events

To test participant's behavior, a limited event zone had to be created, where the monitoring of a participant can be achieved in a well-controlled environment. These zones are the core of the possible measurements in the presented toolbox. Simply put, the event system is realized by a combination of several trigger components. These independent triggers are activated when the participant enters the start trigger (Fig.3: green gate). The event zone is restricted within "boundary" triggers (Fig.3: yellow boxes). These triggers get activated on contact, which is considered a participant's failure. Contact with the event triggers leads to a black screen followed by a respawn of the car at a point after the event (Fig.3: pink box) and giving back the car's control. The successful scenario is when the participant enters the end trigger (Fig.3: red gate) without crashing, i.e., making contact with the "boundary" triggers. All critical events can be adjusted at will, and also a prefabricated file is stored in the repo to create new events. The triggers are all visible in editor mode but invisible to the participant.

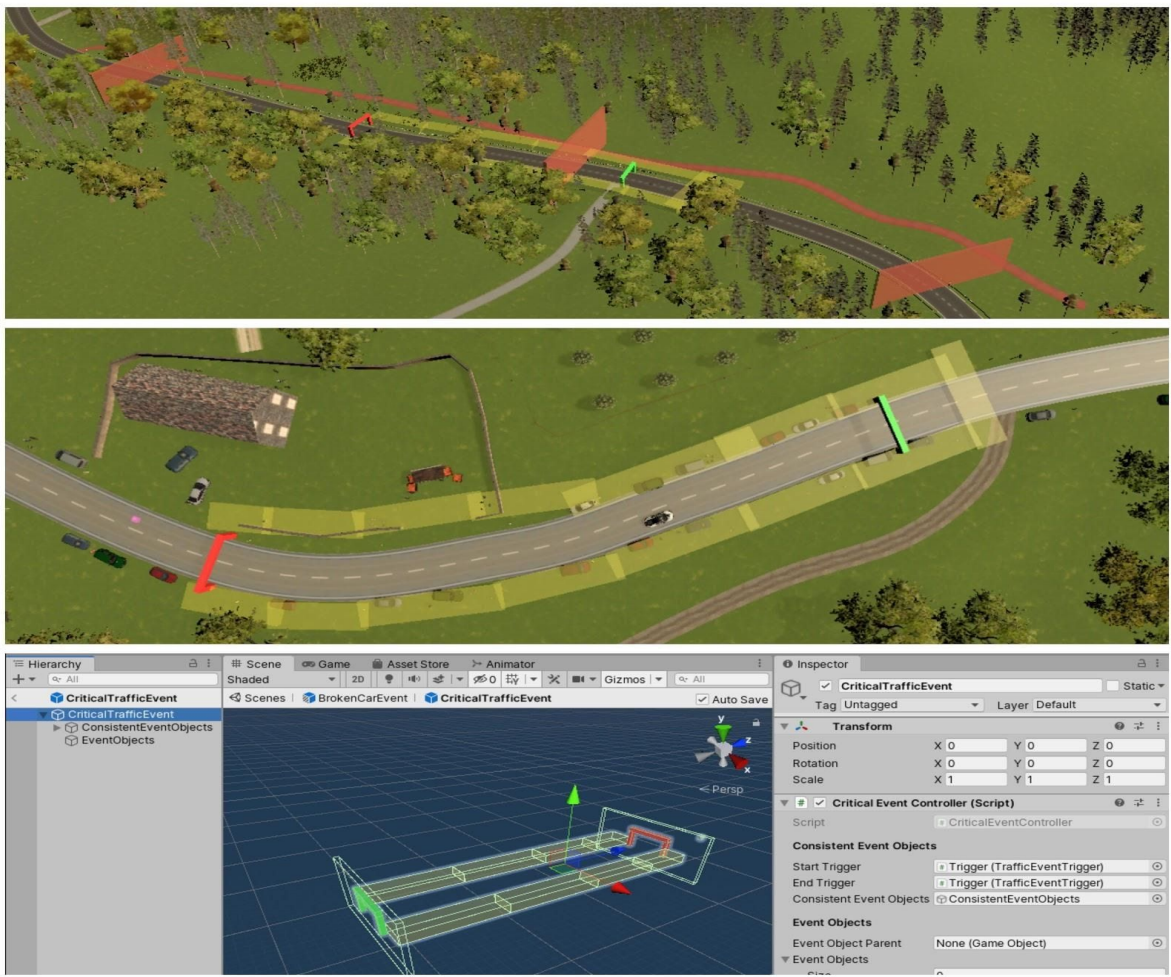

Fig.3: Critical traffic event prefab and its implementation

*2.4. Cars and Traffic Behaviour*

Within the event zones, dynamic objects, such as other road users, are needed to create realistic traffic scenarios. The repository presented here contains various animated pedestrians, animals, and cars to create a broad range of critical situations. All cars used are based on the Unity wheel collider systems of the Unity3D physics engine. In the Car Core Module, user input is translated towards the motor control of the participant's car. The input consists of the motor torque, brake torque, and steering, which are applied to the wheels. This functionality is called AI Control. It allows a seamless transition from automated to manual driving when activated. To facilitate realistic traffic behavior, an additional AI module enables cars to follow predefined paths. Paths followed by AI Cars and walking pedestrians were defined by mathematical bezier curve paths [29], which were realized by a creation tool from previous westdrive projects. Speed limit triggers inside the scene manipulate the AI's aimed speed, which is handling the input propagated to the Car Core Module. Another module of the Car AI allows the AI cars to keep a distance from each other. The goal is to create an easily configurable and interchangeable traffic AI for multiple study designs. With these measures, we maximize the realism of the car physics and traffic simulation while ensuring easy adjustments.

*2.5. Experiment management*

Data sampling, dynamic objects, and driving functionalities within the event zones are controlled by a system of experiment managers that handles scene relevant information and settings shortly before and during the real experiment phase. It handles different camera settings, the information given by triggers inside the scene, and the participants' respawn in case of failure. Before the experiment starts, initial adjustments start the experiment. These adjustments configure the experiment to the participant and include the eye calibration, eye validation, seat calibration, and a test scene.

Fig.4: Scheme of the LoopAR functionalities and components illustrating the interaction of the different services and

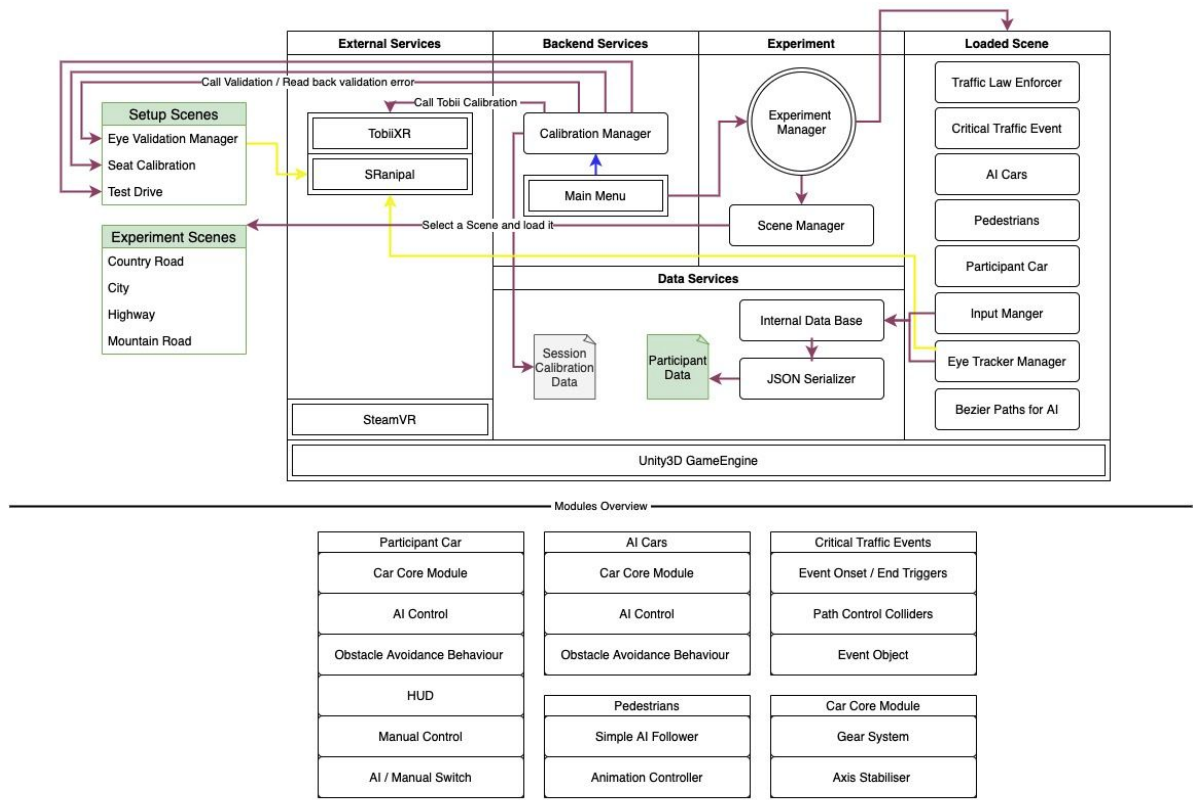

manager scripts within the Unity environment.

The eye-tracking component in this setup comprises an eye-tracking calibration, validation, and online gaze ray-casting, which can record necessary gaze data during the experiment. The component is built for the Tobii HTC Vive Pro Eye device but is intended to keep the VR component interchangeable. Currently, it is intended as a simple connector to tap into SRanipal and the Tobii XR SDK (see Fig. 4). The eye calibration is performed with the built-in Tobii eye calibration tool. The validation is set in the corresponding validation scene, which provides a simple scenario with a fixation cross. Validation fails if the validation error angles exceed an error angle of 1° or the head was moved by 2" from the fixation cross. During the experiment, the eye orientation, position, and collider hits will be stored with a calculated gaze ray of both eyes. Currently, it is set to receive information about any object inside these rays to prevent the loss of viable information by objects covering each other.

Additionally to the eye-tracking data, input data of the participant as well as scene relevant information like the number of failed critical traffic events will be saved using generic data structures and Microsoft Linq, serialized into JavaScript Object Notation (JSON) and saved with a unique ID at the end of each scene. The generic data structure used in the project ensures flexibility, as different data types can be added or removed from the serialization component. The generic data structure used in the project ensures flexibility, as different data types can be added or removed from the serialization component. This approach guarantees the highest compatibility with different analysis platforms such as R or Python for the data gathered with LoopAR.

Data saving and experimental setup aim for a stable and high frame rate, which also avoids motion sickness for subjects. The desired optimum for the experiments is a stable frame rate matching the fixed rate of 90Hz used by the manufacturers HTC and Oculus. Our current frame rate in the different scenes yields an average of 88 samples per second in our test setup, matching the maximum sampling rate of the HTC Vive with 90fps.

## 3. Hardware requirements

The setup used and presented here is thought to be a cost-efficient and very mobile replacement for maintenance-intensive, rigid, and expensive driving simulators for studies on human behavior in the context of self-driving cars. A key advantage is freedom regarding the selected components. The only requirement for operation is granting the computing power for the entire system, which consists of a core setup only of a computer, a head-mounted display, and a steering wheel.

As a virtual reality device, we use the HTC Vive Pro Eye with an integrated Tobii Eye Tracker. It is a cable bound head-mounted-display that enables the participant to transfer movements into virtual reality. Although we are using the Vive Pro exclusively at our department, the LoopAR experiment is not dependent on this specific VR device.

Used components of the setup with 90fps sampling and display:

| GPU | Nvidia GeForce RTX 2080, equivalent or better |
|---|---|
| CPU | Intel(R) Xenon RE5-1607 v4, equivalent or better |
| RAM | 32 GB |
| Video Output | HDMI 1.4, DisplayPort 1.2 or newer. USB Port: 1x USB 2.0 or better port |

| Operating System | Windows 8.1 or later, Windows 10 |
|---|---|
| VR HMD | Vive Pro -Eye with built-in Tobii Eye Tracker |
| Steering Wheel | Game-ready Fanatec CSL Elite Steering Wheel and pedals |

**4. Discussion**

In the presented paper, we developed LoopAR as a modular toolkit to test a take over of control in critical traffic situations from automated cars to human drivers by combining VR and eye-tracking in an interactive and immersive scenario. In its current state and design, it provides a promising new low-cost and mobile setup to conduct studies that were traditionally only done in stationary simulators.

Also, this project quickly enables users to adapt the current code and 3D environments to create experiments in VR. Due to the open-access functionalities of the project, it can have a significant influence on the way future VR studies are conducted. With the newly implemented code, it is not only possible to simulate a large and highly realistic VR environment, but it is also possible to create a broad range of applications in VR research that is not only bound to HMI investigations. A large part of the assets used is from Unity's asset store and the 3D platforms Sketchfab and Turbosquid. Each of the four scenes can be customized anytime. It is, therefore, possible to change the number, size, and shape of all objects in each scene.

All of the functionalities above and assets presented here are all under constant improvement. By writing, five new projects arise from the presented toolkit, which will also develop new assets and features implemented into the toolkit later on. The authors want to emphasize the modularity and adaptability of this VR toolkit. We are happy to present LoopAR, and we are already looking forward to the many extensions of the WestDrive project.

**5. Conclusions**

This article describes a new virtual reality tool kit for unity applications investigating the human-machine interaction in highly automated driving developed by us. The presented setup is thought to be a mobile, cost-efficient, and highly adaptable alternative to chassis simulators that closely monitor the participants. Particularly noteworthy is not only the drastic reduction in costs but also the adaptability of the software as well as the used hardware. All components are fully upgradable, in case there are better products in terms of image quality or computing power. LoopAR allows interested researchers to conduct a variety of virtual reality experiments without having to create the needed environment or functionalities themselves. For this, we have provided an area of almost 25km² based on OSM data. The toolkit presented here also includes all the necessary assets and basic prefabs to quickly and precisely create a wide variety of virtual environments. Additionally, the LoopAR toolkit includes components of the experimental procedure and data storage.

**Data availability statement**

The datasets for the asset foundation and scripts can be found in the WestDrive repository
**https://github.com/Westdrive-Workgroup/MotorCity-Core**

The dataset for the LoopAR project is accessible at:
**https://github.com/Westdrive-Workgroup/LoopAR**

**Supplementary Materials:**
Unity® 3D: www.Unity3d.com
Online Character animation: www.mixamo.com
Adobe Fuse CC: www.adobe.com/products/fuse.html
Blender 2.81: www.blender.org

**Author Contributions:** FNN and MAW wrote this paper. Both authors designed the project. SK, PK, and GP supervised the LoopAR project. NM developed major parts of the AI and Functional Modules and User Interfaces. JP realized scene building and the HUD functionalities. LK designed the mountain road scene and provided performant assets. LT developed large parts of the software architecture and acted as a software engineer for the project's functional compartments. AH designed the highway scene and provided additional assets. LMK designed the country road scene, provided assets, and contributed to HUD related literature background. PS was involved in designing the city scene, managing and creating assets. FN developed and designed the TestDrive Scene.

**Funding:** This work is funded by the University of Osnabrück in cooperation with the graduate college "Vertrauen und Akzeptanz in erweiterten und virtuellen Arbeitswelten" (FNN), as well as from the GMH foundation https://www.stiftung-stahlwerk.de/home/ (MAW)

**Conflicts of Interest:** The authors declare that the research was conducted in the absence of any commercial or financial relationships that could be construed as a potential conflict of interest.
.